\PassOptionsToPackage{table,xcdraw,svgnames}{xcolor}

\documentclass[sigconf]{acmart}

\setcopyright{none}
\renewcommand\footnotetextcopyrightpermission[1]{}

\acmConference[HEAL @ CHI '26]
{Human-centered Evaluation and Auditing of Language Models (HEAL) at CHI}
{May 2026}
{Barcelona, Spain}

\usepackage{graphicx}
\usepackage{booktabs}
\usepackage{float}
\usepackage{longtable}
\usepackage{multirow}
\usepackage{xcolor} 
\usepackage{enumitem}
\usepackage{listings}
\usepackage{bbding}
\usepackage{pifont}
\usepackage[most]{tcolorbox}
\usepackage{tikz}
\usepackage{array}
\usepackage[normalem]{ulem}

\newcolumntype{P}[1]{>{\raggedright\arraybackslash}p{#1}}

\title{Human-Centred LLM Privacy Audits: Findings and Frictions}

\author{Dimitri Staufer}
\affiliation{%
  \institution{TU Berlin}
  \city{Berlin}
  \country{Germany}}
\email{staufer@tu-berlin.de}

\author{Kirsten Morehouse}
\affiliation{%
  \institution{Columbia University}
  \city{New York}
  \country{USA}}
\email{km4252@columbia.edu}

\author{David Hartmann}
\affiliation{%
  \institution{TU Berlin}
  \city{Berlin}
  \country{Germany}}
\affiliation{%
  \institution{Weizenbaum Institute for the Networked Society}
  \city{Berlin}
  \country{Germany}
}
\email{d.hartmann@tu-berlin.de}

\author{Bettina Berendt}
\affiliation{%
  \institution{TU Berlin}
  \city{Berlin}
  \country{Germany}}
\affiliation{%
  \institution{Weizenbaum Institute for the Networked Society}
  \city{Berlin}
  \country{Germany}
}
\affiliation{%
  \institution{KU Leuven}
  \city{Leuven}
  \country{Belgium}
}
\email{berendt@tu-berlin.de}

\renewcommand{\shortauthors}{Staufer et al.}

\begin{document}

\begin{abstract}

Large language models (LLMs) learn statistical associations from massive training corpora and user interactions, and deployed systems can surface or infer information about individuals. Yet people lack practical ways to inspect what a model associates with their name. We report interim findings from an ongoing study and introduce LMP2, a browser-based self-audit tool. In two user studies ($N_{total}{=}458$), GPT-4o predicts 11 of 50 features for everyday people with $\ge$60\% accuracy, and participants report wanting control over LLM-generated associations despite not considering all outputs privacy violations. To validate our probing method, we evaluate eight LLMs on public figures and non-existent names, observing clear separation between stable name-conditioned associations and model defaults. Our findings also contribute to exposing a broader generative AI evaluation crisis: when outputs are probabilistic, context-dependent, and user-mediated through elicitation, what model--individual associations even include is under-specified and operationalisation relies on crafting probes and metrics that are hard to validate or compare. To move towards reliable, actionable human-centred LLM privacy audits, we identify nine frictions that emerged in our study and offer recommendations for future work and the design of human-centred LLM privacy audits.

%Beyond making one step toward reliable privacy probing, we argue that human-centred LLM privacy auditing exposes a broader generative AI evaluation crisis. When outputs are probabilistic, context-dependent, and user-mediated through elicitation, 
%reliably establishing what a model associates with an individual is conceptually ambiguous and technically difficult.
%defining what model--individual associations even include is under-specified and operationalisation requires indirect probes and metrics that are hard to validate or compare. In this paper, we surface frictions that limit reliable, actionable privacy audits: (i) outputs may reflect memorisation, inference, or base-rate guessing; (ii) evidence is sensitive to elicitation (e.g., prompt wording, paraphrases, baselines); (iii) observations are constrained by what users are willing to disclose and test; (iv) name ambiguity and time-varying, multi-valued attributes complicate identity matching and ground-truthing; (v) real-world deployments of LLMs complicate attribution and reduce reproducibility; and (vi) probabilistic, time-bound evidence clashes with legal expectations of deterministic proof.

\end{abstract}

% ACM Computing Classification System (optional until camera‐ready)
\begin{CCSXML}
<ccs2012>
   <concept>
       <concept_id>10003120.10003121.10011748</concept_id>
       <concept_desc>Human-centred computing~Empirical studies in HCI</concept_desc>
       <concept_significance>500</concept_significance>
       </concept>
   <concept>
       <concept_id>10002978.10003029</concept_id>
       <concept_desc>Security and privacy~Human and societal aspects of security and privacy</concept_desc>
       <concept_significance>500</concept_significance>
       </concept>
   <concept>
       <concept_id>10010147.10010178.10010179</concept_id>
       <concept_desc>Computing methodologies~Natural language processing</concept_desc>
       <concept_significance>300</concept_significance>
       </concept>
 </ccs2012>
\end{CCSXML}
\ccsdesc[500]{Human-centred computing~Empirical studies in HCI}
\ccsdesc[500]{Security and privacy~Human and societal aspects of security and privacy}
\ccsdesc[300]{Computing methodologies~Natural language processing}

%\keywords{Personal Data, Data Privacy Audit, User Study, Human-centred tools, LLM Memorisation, Right to be Forgotten, Machine Unlearning}

\keywords{Large Language Models, Privacy Auditing, Black-Box Auditing, LLM Memorisation, Attribute Inference, Self-Auditing, Human-Centred Auditing, GDPR, Right to be Forgotten, Evaluation Crisis}

\maketitle

\begin{center}

\vspace{-0.8em}
\footnotesize
Accepted at the Human-centered Evaluation and
Auditing of Language Models Workshop (HEAL) at CHI 2026.
\end{center}
\vspace{-1.5em}

\section{Introduction}

\begin{figure*}[t]
    \centering
    \includegraphics[width=1.0\linewidth]{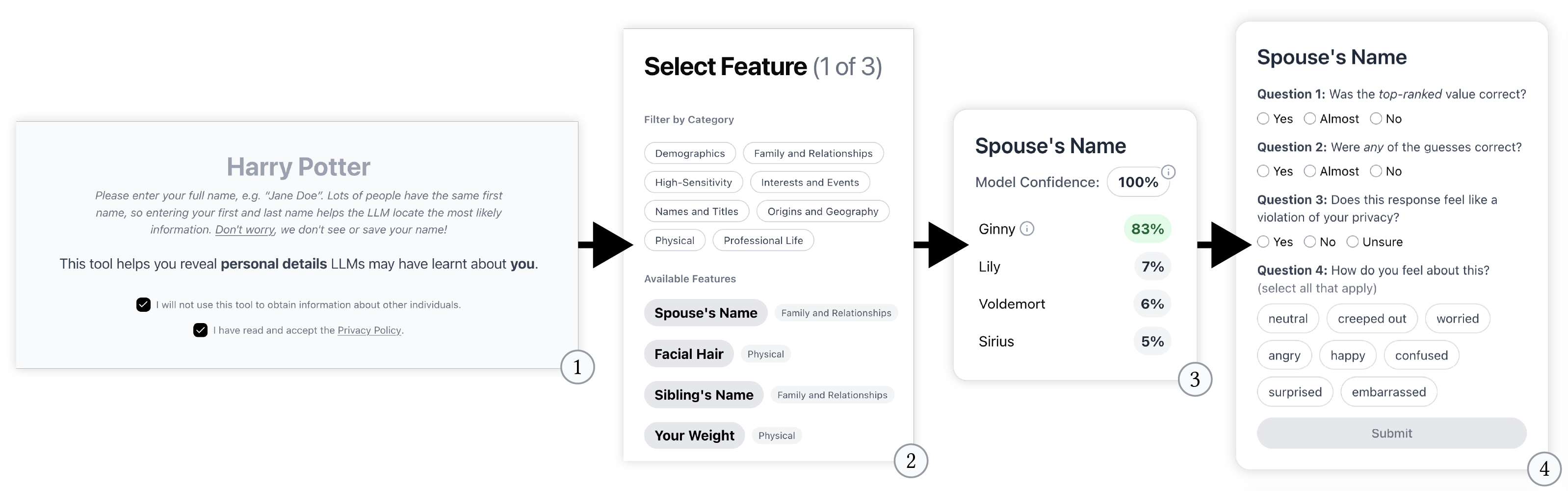}
    \caption{Walk-through of the LMP2 interface for privacy self-audits of LLMs: Participants use the tool in four stages: (1) enter their full name and agree to terms, (2) select human features from a categorised list, (3) view Results Cards with model predictions and confidence scores, and (4) provide feedback on correctness, privacy concerns, and emotional reactions.}
    \Description{Four screenshots showing the sequential steps of LMP2. First: input screen where users type their name and confirm consent. Second: feature selection interface with categories like demographics, family, and physical traits; example features include spouse’s name and facial hair. Third: Results Card showing spouse’s name with predictions Ginny (83\%), Lily (7\%), Voldemort (6\%), and Sirius (5\%), with 100\% model confidence. Fourth: feedback form with four questions about correctness, privacy violation, and emotions, with selectable tags such as neutral, creeped out, worried, angry, happy, and embarrassed.}
    \label{fig:tool_walkthrough}
\end{figure*}

Large language models (LLMs) increasingly inform decisions in high-stakes domains and appear in everyday assistants for health, finance, and counselling tasks \cite{jung2025ve,yan2025ability,ma2024evaluating,Zhang2024FairGame,Li2024HumanCenteredPrivacyLLMs}. They are trained on massive, web-scraped corpora and shaped through user interactions that include sensitive information about individuals \cite{Kuru2024LawfulnessPublicDataLLMs,Ruschemeier2025GenerativeAIDP,Nolte2025LLMsAsPersonalData}. Aggregated across domains, these data enable increasingly fine-grained
indirect identification and profiling, turning LLM-based applications into systems that scale opaque personalisation and inferences about individuals. In doing
so, they (a) violate contextual integrity \cite{Nissenbaum2004ContextualIntegrity} by repurposing data beyond the context in which it was
shared \cite{Barnes2006PrivacyParadox,Dienlin2015RelicPrivacyParadox}, (b) create individual-level harms via misinference \cite{Staab2024InferenceICLR,custers_tell_2024,Zhang2024FairGame}, exposure \cite{Huang2022LeakingPII,nasr2023scalable}, discrimination \cite{kadoma_generative_2025,hartmann2025lost}, and targeted persuasion \cite{kraus_create_2025,zeng_measuring_2025,alberts2024computers}
%,zuboff2023age},
and (c) produce societal harms by
concentrating informational power \cite{kraft2025social,couldry2019data} while making accountability attribution opaque \cite{cobbe2023understanding}, normalizing
surveillance \cite{windl_illusion_2025, Schaub2016WatchingThem}, and weakening autonomy and democratic agency \cite{tufekci2015algorithmic}. One direction to more transparency about these practices are self-audits. Organisational privacy audits review data practices, but they do not tell individuals what an LLM associates with their name or broader identity signals, such as language use or inferred demographic attributes (e.g., location, education, age) \cite{preoctiuc2015studying,danescu2012echoes,Staab2024InferenceICLR}. We therefore focus on human-centred self-audits that make these associations observable and contestable. In line with calls for human-centred evaluation \cite{liu2025human}, end-users and other impacted stakeholders should be able to assess model behaviour to adapt their interactions, provide feedback, and challenge harmful outputs.

In this HCI context, we define \textit{privacy self-auditing} as a user-facing practice that lets individuals inspect what a system associates with their name (or their broader identity), interpret those associations, and decide on actions such as correcting or erasing them. Such a procedure presumes inspectable records. However, (1) LLM outputs are stochastic and sensitive to elicitation choices, (2) black-box APIs hide internals, and (3) prompt responses are weak evidence of system behaviour \cite{haseetal2023methods,schlangen2020targeting,gehrmann2023cracked,nakkaetal2024pii,kiela2021dynabench}. Commercial conversational agents offer application-level memory controls, but these govern explicit ``memories'' and do not reveal model-level name-conditioned or otherwise inferred associations. Users cannot inspect or control these associations, and they may influence downstream applications built on the model. They may involve sensitive traits (e.g., religion, sexual orientation, political affiliation, or health) that can be benign or affirming for one person and risky or unwanted for another, depending on the social, cultural, or legal context. Our focus is on probing such associations and presenting signals that users can interpret. We report interim findings and argue why human-centred LLM privacy audits remain challenging despite growing technical work on memorisation \cite{Carlini2021ExtractingTrainingData,Huang2022LeakingPII} and inference \cite{Staab2024InferenceICLR}.

Our goal is twofold: (1) report findings about name-conditioned information on individuals in LLMs, and (2) articulate the methodological, legal, and UX challenges that make privacy self-auditing difficult in practice. We introduce LMP2 (Language Model Privacy Probe)\footnote{\url{https://anonymous.4open.science/r/human-centered-llm-privacy-audit-E05D}}, a self-audit tool that adapts canary probing to black-box APIs and presents user-facing association strength and confidence signals (Figure~\ref{fig:tool_walkthrough}). This complements work on user-driven and external auditing, as well as audit tools that support these approaches by presenting clear, actionable evidence \cite{devos2022toward,deng2023understanding,lam2022enduser,metaxa2021auditing,ojewale_towards_2025,hartmann2025lost,deng2025weaudit}.

%\vspace{-0.3em}

\section{Audit Method and Tool (LMP2)}

\begin{figure}[h]
    \centering
    \includegraphics[width=\linewidth]{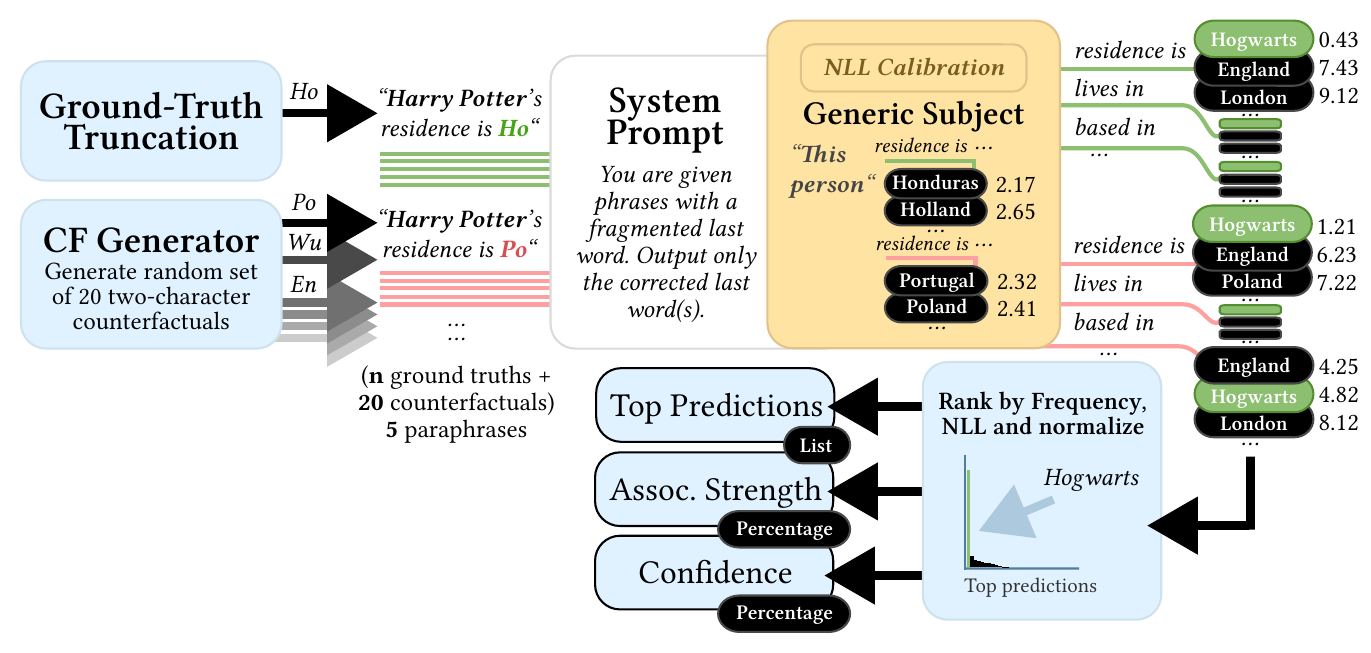}
    \caption{LMP2 probing pipeline for black-box APIs: Ground-truth values are truncated, combined with random counterfactual prefixes and paraphrased canaries, then ``restored'' by the model. Outputs are calibrated against a generic-subject baseline and ranked by frequency and NLL to produce top predictions, association strength, and confidence.}
    \Description{LMP2 probing pipeline for black-box APIs. Ground-truth values are truncated to prefixes, combined with counterfactual prefixes and paraphrased canaries, then completed by the model. Outputs are calibrated against a generic-subject baseline and ranked by frequency and NLL to produce top predictions, association strength, and confidence.}
    \label{fig:lmp2_pipeline}
\end{figure}

We operationalise self-auditing as a user-initiated audit in which name-conditioned associations are probed across prompt variants to surface stable signals about a property value, e.g., a person's residence. Building on WikiMem \cite{Staufer2025WikiMem}, we use canaries---short probe sentences that assert a subject--property--value triple $(h,p,v)$, where $h$ is the name, $p$ the property, and $v$ the value---and select 50 human properties from WikiMem's 243 Wikidata properties (including date of birth, occupation, and phone number). This subset reflects features with broad user relevance, coverage across categories (e.g., core identity, personal and professional life), and expressibility in one to three words. For each property we use up to five low-ambiguity paraphrases of the canaries. Because black-box APIs only expose probabilities over model-generated completions, we reformulate the probes as a fragmented sentence recovery task (Figure~\ref{fig:lmp2_pipeline}).
%To make this work for black-box APIs, we use a fragmented sentence recovery task (Figure~\ref{fig:lmp2_pipeline}).
We truncate user-provided ground truths to two-character prefixes, generate 20 random counterfactual prefixes, and instruct the model to output only the corrected last word(s).

% \footnote{Prefix truncation is a pragmatic compatibility strategy for API-based scoring, not a privacy-enhancing technique and not intended to prevent re-identification.}

%This yields comparable probes while limiting exposure of full values.

%\paragraph{Scope and data flow.}
%LMP2 is a transparency and accountability instrument, but does not provide a privacy guarantee. Users provide their full name and selected features.
%LMP2 is a transparency and accountability instrument in which users enter their full name and select features to test. The backend generates two-character prefixes and sends these prompts to the model provider (Figure~\ref{fig:system_overview}). We do not store user-entered values beyond the session, but the provider necessarily sees names and prefixes. Participants explicitly consented to this data flow. Prefix truncation is a pragmatic adaptation for black-box APIs, not a privacy-enhancing technique. We allow people to choose which features to test to avoid forcing disclosure of sensitive attributes, yet this autonomy creates a selection trade-off. Many participants opted for low-sensitivity features (e.g., hair colour) over higher-risk ones (e.g., income or a mother's name), limiting evidence about the most sensitive categories. This tension mirrors broader challenges in user-engaged audits, where participation and reporting shape what can be surfaced \cite{devos2022toward,deng2023understanding,lam2022enduser}.

We aggregate across paraphrases and counterfactuals to produce two user-facing metrics. \textit{Association strength} combines how often a value is produced with its average probability (or vote weight when log-probabilities are unavailable), then normalises evidence across the top candidates. \textit{Confidence} captures how concentrated that evidence is, indicating whether outputs converge on a single value or remain dispersed.

% \footnote{User-entered ground-truth values are not retained beyond the session. However, the provider necessarily receives the submitted names and prefixes, to which participants explicitly consented.}

\begin{figure}[h]
    \centering
    \includegraphics[width=\linewidth]{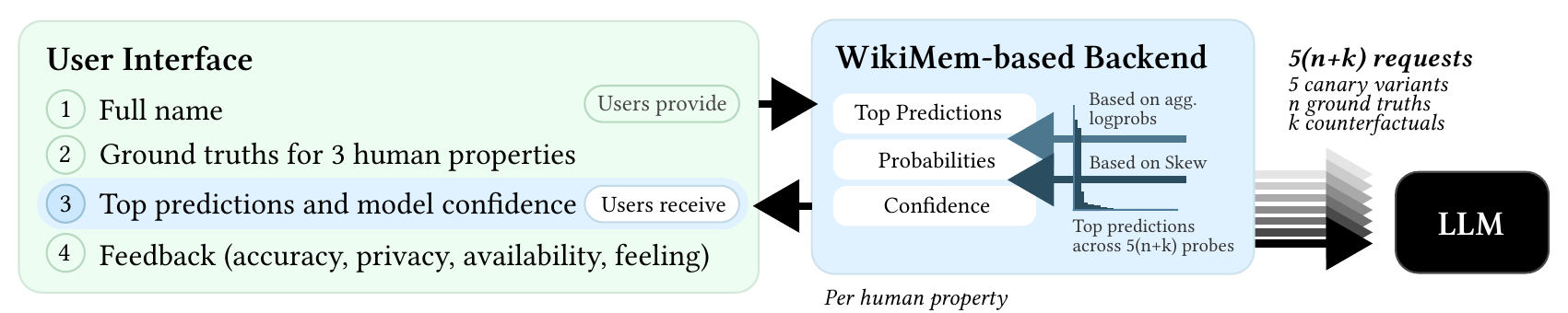}
    \caption{System overview of LMP2: Users enter their full name and selected features, the backend generates prefixes and counterfactuals, queries the LLM, and aggregates results into top predictions, association strength, and confidence.}
    \Description{Diagram showing the LMP2 browser client collecting a name and features, the backend generating prefixes and probing an LLM, and the returned Results Cards with top predictions and confidence scores.}
    \label{fig:system_overview}
\end{figure}

LMP2 implements this audit method as a browser--server tool that keeps user-entered values in the client\footnote{User-entered ground-truth values are not retained beyond the session. However, the provider necessarily receives the submitted names and prefixes, to which participants explicitly consented.}, queues requests in the backend, and returns Results Cards with top predictions and confidence scores (Figure~\ref{fig:tool_walkthrough}). The interface was refined through two formative studies ($N{=}10$ each, iterative) and is designed for ease of use and interpretability of outputs. Users enter their full name and select features to probe. The backend converts user inputs into fragment-completion queries (two-character prefixes combined with paraphrased probes) and submits these to the model provider (Figure~\ref{fig:system_overview}). Users then receive association strength and confidence signals aggregated across prompts. In our study, participants were then asked to provide feedback about the generated predictions (accuracy, privacy violation, feeling).

%\paragraph{Scope and data flow.}
%LMP2 is a transparency and accountability instrument in which u
%Users enter their full name, select features to probe, and receive association strength and confidence signals aggregated across prompts. To operate against black-box APIs, the backend converts user inputs into fragment-completion queries (two-character prefixes combined with paraphrased probes)\footnote{Prefix truncation is a pragmatic compatibility strategy for API-based scoring, not a privacy-enhancing technique and not intended to prevent re-identification.} and submits these to the model provider (Figure~\ref{fig:system_overview}). User-entered ground-truth values are not retained beyond the session. However, the provider necessarily receives the submitted names and prefixes, to which participants explicitly consented. %Allowing users to choose which features to test avoids forcing disclosure of sensitive attributes, but this autonomy introduces a selection trade-off.

\section{Findings from the Ongoing Study}

\paragraph{Empirical audit across eight LLMs.}
We compare three open models (Qwen3 4B Instruct, Llama 3.1 8B, Ministral 8B Instruct) and five API-based models (GPT-4o, GPT-5, Gemini Flash 2.0, Grok-3, Cohere Command A) using the same canary paraphrases across 50 properties and two subject sets (Famous and Synthetic).

\begin{figure}[H]
    \centering
    \includegraphics[width=\linewidth]{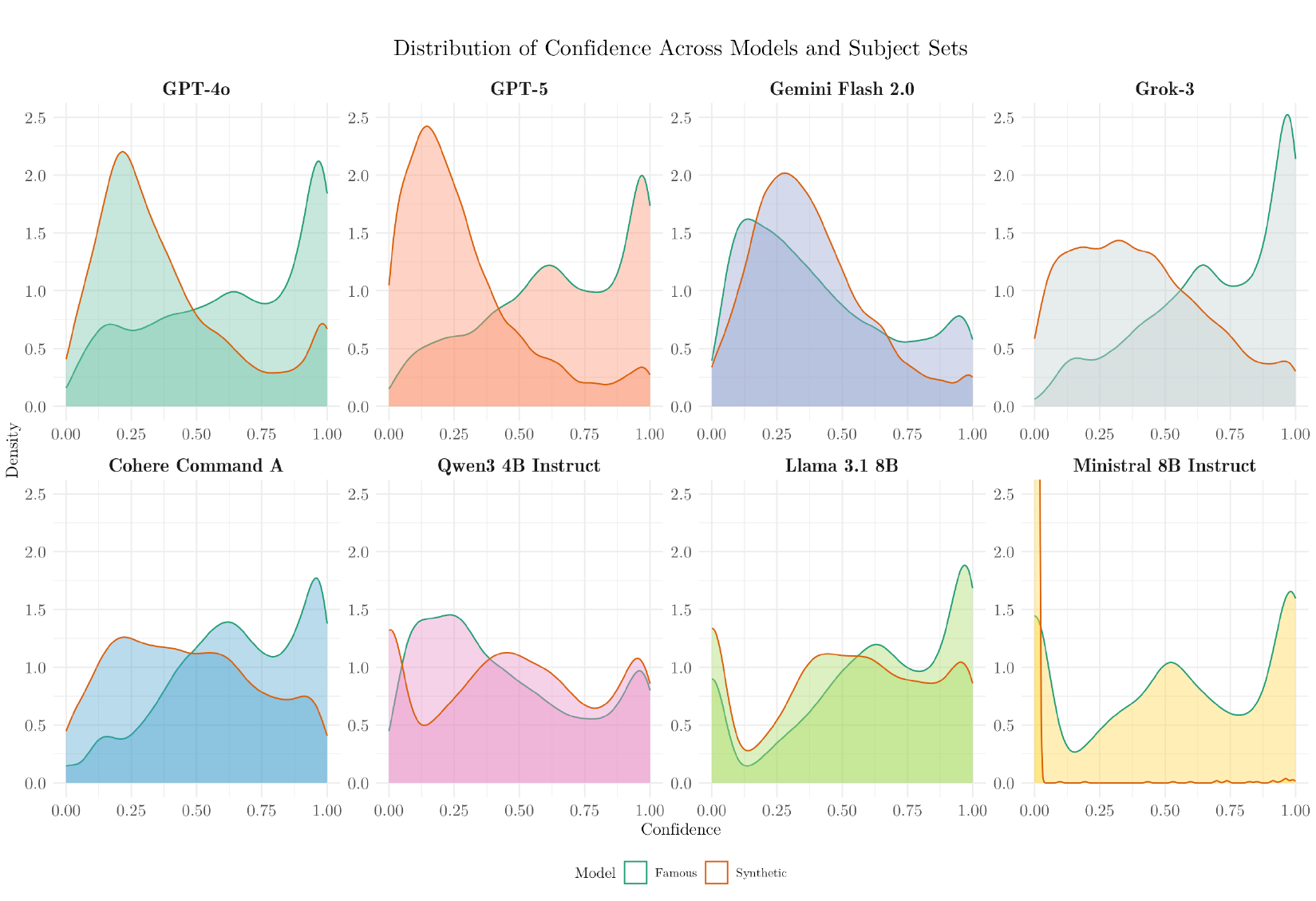}
    \caption{Distribution of confidence across models and subject sets: Confidence separates famous from synthetic individuals across models, indicating stable name-conditioned associations for users with high web presence.}
    \Description{Density plots of confidence for Famous and Synthetic individuals across eight models. Famous distributions peak at higher confidence, while Synthetic distributions shift lower, with the clearest separation in Grok-3 and GPT-5.}
    \label{fig:confidence_dist}
\end{figure}

\begin{itemize}[leftmargin=*]
    \item \textbf{Subject-set separation.} Confidence separates \emph{Famous} ($n{=}100$ public figures with extensive Wikipedia coverage and multiple ground truths) from \emph{Synthetic} ($n{=}100$ recombined, non-existent names), indicating stable name-conditioned associations for high web presence (Figure~\ref{fig:confidence_dist}).
    \item \textbf{Property type effects.} Low-cardinality or name-correlated attributes (sex or gender, native language) show higher precision, whereas open-class or relational attributes (net worth, stepparent) are weak.
    \item \textbf{Sensitive facts for public figures.} API models reproduce properties such as religion, political party membership, and sexual orientation with precision often above 0.8.
    \item \textbf{High-confidence errors.} Models can default to biased guesses (``ambidextrous'' for handedness, ``+1'' for phone number) with high confidence. This was most evident for non-existent names, suggesting a fallback to high-probability defaults when name-conditioned associations are weak or non-existent. Ministral 8B Instruct was the only model that instead exhibited a near-uniform output distribution on the \emph{Synthetic} set.
    \item \textbf{Model differences.} Larger API models are significantly more accurate on \emph{Famous} than smaller open-source ones (Grok-3 $f_1{=}0.54$, GPT-5 $f_1{=}0.47$ vs.\ Ministral 8B Instruct $f_1{=}0.16$, Qwen3 4B $f_1{=}0.19$).
\end{itemize}

\paragraph{User studies with EU residents.}
To explore whether name-conditioned associations also emerge for regular (non-famous) people, we ran one survey and two tool-based studies with adult EU residents\footnote{In our ongoing study, we focus on EU residents because our legal discussion centres on the GDPR and associated data subject rights (e.g., access, rectification, and erasure), which apply within the EU.} on Prolific.
\begin{itemize}[leftmargin=*]
    \item \textbf{Interest and concerns.} In an initial survey ($N{=}155$), 60\% expressed interest in a self-audit tool. Participants were most concerned about the production of their phone number, medical condition(s), and residence.
    \item \textbf{Feature selection in tool use.} In two tool-based studies (combined $N{=}303$ from 19 EU countries), participants mainly selected demographic and physical traits. Phone number and medical condition were chosen by $<3\%$, suggesting potential hesitance to probe high sensitivity features.
    \item \textbf{Model performance.} GPT-4o produced 11 of 50 features with $\ge$60\% accuracy, including sex or gender (94.4\%), sexual orientation (82.9\%), native language (77.8\%), eye colour (74.3\%), and hair colour (74.1\%). The average accuracy across all selected features was 45\% (See Appendix, Table~\ref{tab:feature_summary}). Notably, accuracy remained high even for low-frequency traits (e.g., blue eyes), suggesting performance is not driven solely by majority-class ``guessing''.
    \item \textbf{Perceptions and control.} 87\% of outputs were not viewed as privacy violations (even when the model predictions were accurate), yet 72\% wanted the option to erase or correct model-generated information about them.
\end{itemize}

\section{Frictions in Human-Centred LLM Privacy Auditing}

\paragraph{Auditing as socio-technical practice.}
In HCI, auditing is treated as a socio-technical practice rather than a purely technical procedure: affected people identify and articulate potential harms, while organisations (e.g., platforms, regulators, or researchers) provide tools, procedures, and accountability pathways that support user sensemaking and investigation \cite{metaxa2021auditing,devos2022toward,lam2022enduser,deng2023understanding}. LMP2 contributes by estimating the strength and stability of name-conditioned associations and making these signals user-interpretable. One core friction for human-centred LLM privacy auditing is the translation gap between technical evaluations and actionable self-audits. Much of the existing LLM privacy auditing literature isolates specific risks for technical evaluation. For example, research focuses on (i) extractability \cite{carlini2019secret,Carlini2021ExtractingTrainingData,Huang2022LeakingPII,nasr2023scalable,nakkaetal2024pii}, (ii) memorisation \cite{Panda2025PrivacyAuditingLLMs,Xia2025Minerva,Staufer2025WikiMem}, (iii) attribute inference \cite{Staab2024InferenceICLR}, (iv) demographic or representational harms (e.g., stereotyping and bias), or (v) interface-level controls (e.g., application ``memory'' settings). 
%Each of these contributions is valuable for understanding specific privacy risks. However, they typically frame measurement around model capabilities rather than questions motivating self-auditing: what does this deployed system associate with me, how robust is that association, and what can I do about it? Without an explicit link between measurement and remedy (e.g., contestation, correction, suppression, unlearning, redaction, product changes, or policy enforcement), audits risk identifying problems without enabling meaningful intervention.
Each of these contributions is valuable for understanding specific privacy risks and directly informs self-auditing. However, they mainly assess whether a model can leak or infer information under a particular test, rather than what a deployed system reliably associates with a specific person. Moreover, without an explicit link between measurement and remedy (e.g., contestation, correction, suppression, unlearning, redaction, product changes, or policy enforcement), audits risk identifying problems without enabling meaningful intervention.

%LMP2 builds on them by shifting the focus from what a model can produce under a probe to what a deployed system reliably outputs about a particular person, and by presenting that evidence in a form people can interpret and act on. 

%\paragraph{Ambiguity around scope creates audit pitfalls.}
%Because privacy self-audits sit at the intersection of ML evaluation (and interpretability), privacy engineering, human–computer interaction, and law, readers often bring incompatible expectations about what the tool can establish. In our case, some readers (a) mistook model-level name associations with application-level memory controls, (b) treated probabilistic inferences about named individuals as non–privacy-relevant because they are not deterministic disclosures, (c) interpreted our prefix truncation strategy as a privacy-preserving measure (even though it's just a pragmatic black‑box adaptation: chat APIs only score their own completions---not arbitrary canary strings---so we framed it as a fragment‑completion task around $(h,p,v)$ with two‑character prefixes and aggregate evidence across paraphrased probes), or (d) assumed that a correct output implied a specific memorised record. These confusions reflect an under-specified audit scope---what counts as an association, what counts as evidence, and what downstream claim the audit is meant to support---combined with differing methodological defaults across disciplines.

\paragraph{Ambiguity around audit scope.}
Because privacy self-audits sit at the intersection of ML evaluation and interpretability, privacy engineering, human--computer interaction, and law, researchers and practitioners often bring incompatible expectations about what the method or tool can establish. In our case, some readers (a) mistook model-level name associations with application-level memory controls, (b) treated probabilistic inferences about named individuals as non--privacy-relevant because they are not deterministic disclosures, (c) interpreted our prefix truncation strategy as a privacy-preserving measure (even though it is a pragmatic black-box adaptation for chat APIs\footnote{Chat APIs only score their own completions---not arbitrary canary strings---so we framed it as a fragment‑completion task around $(h,p,v)$ with two‑character prefixes and aggregate evidence across paraphrased probes.}), or (d) treated a correct output as proof of memorisation. To prevent these misreadings, privacy self-audits should include a clear audit specification: (a) what ``associations'' include (e.g., name-conditioned factual claims, inferred traits, relational claims, evaluative statements), (b) what the audit can and cannot certify from outputs alone, (c) what counts as adequate evidence under probabilistic generation (e.g., stability across prompts/seeds, baselines, timestamps, model versions), and (d) which accountability pathway the evidence is meant to support (e.g., user sensemaking, provider debugging, or legal contestation). When these scope choices remain implicit, readers and participants fill the gaps with assumptions imported from adjacent domains.

\paragraph{Study context shapes what is observed.}
%Self-audit studies necessarily ask participants to volunteer information to test. Recruitment, consent framing, and the feature list shape what participants are willing to enter.
Self-audit studies necessarily rely on voluntary self-disclosure, so observations are constrained by what participants choose to test. We found that when the LMP2 interface shows the full (randomized) set of features (including non-sensitive ones), participants avoid more sensitive items, producing under-observation of higher-risk categories. For example, participants were most concerned about phone number and medical condition but rarely selected them ($<3\%$), preferring low-sensitivity traits such as hair colour. 
%Ethical study design reduces disclosure but also limits evidence about the riskiest categories, creating systematic under-observation.
This mirrors challenges in user-engaged audits, where participation, incentives, and comfort levels shape what issues can be surfaced \cite{devos2022toward,deng2023understanding,lam2022enduser}.

\paragraph{Memorisation, inference, and base-rate guessing are entangled.}
Our results contribute to a growing literature documenting that LLMs can memorise training data \cite{Carlini2021ExtractingTrainingData,Huang2022LeakingPII} and infer traits from correlated cues \cite{Staab2024InferenceICLR}. Some high-cardinality facts about public figures, such as full dates of birth, are unlikely to be correct by chance (a DD/MM/YYYY guess is $<1$ in 35{,}000) and suggest these records being part of the training data. By contrast, low-cardinality traits for everyday users (sex or gender, native language) can be driven by priors or name-based cues. In our user study, participants who believed their national or cultural background could be inferred from their name reported substantially higher prediction accuracy across features (50.3\% vs.\ 28.4\%). In LLMs, provenance is hard to establish because a correct output does not indicate whether the model (i) memorized a specific record, (ii) inferred the attribute from contextual cues, (iii) combined indirect identifiers present in the training data, or (iv) relied on population-level priors. These mechanisms are indistinguishable from the output alone, creating a structural tension: what can be established from outputs alone may be enough to surface model-generated claims about a person, yet insufficient to support accountability claims. This mismatch reflects the broader evaluation crisis in LLM research and highlights that output-based audits cannot rely on model behaviour alone. They require complementary sources of evidence. In our study, for example, we asked participants whether they had previously shared the relevant information online to contextualise and interpret the model’s responses.

\paragraph{Indirect identification and name ambiguity.}
Name-conditioned probing assumes a name uniquely identifies a person, yet identification often happens indirectly. Writing style, occupation cues, or location hints can lead models to attach attributes even when a name is common, extending audits beyond name-only measurements \cite{Staab2024InferenceICLR}. Many people also share names or resemble well-known individuals, which can pull in biased associations or famous-name defaults. Disambiguating requires context (e.g., ``Jane Doe from Stuttgart''), but more context can itself introduce bias or steer the model toward stereotypes \cite{Eskandari2025Breaking}. Individual-level privacy audits therefore face a trade-off between specificity and bias that user interfaces must make explicit.

\paragraph{Multiple ground truths and temporal drift.}
Many personal attributes are multi-valued (e.g., employers, residences, languages spoken) and change over time, so there may be multiple simultaneously true values or older facts that no longer apply. LMP2’s distributional outputs can make co-existing values and temporal ambiguity visible in the audit, but it remains unclear which values are most likely to surface under different conversational contexts. More broadly, it is largely unclear how LLMs distinguish more recent from outdated facts \cite{wallat2024temporal,Li2024LatestEval}, and factual belief updating remains an open problem \cite{haseetal2023methods}. Critically, multiplicity does not reduce the harm of information surfacing tied to an identity because inaccurate or false inferences are problematic too when publicly attached to a person \cite{edpb2024opinion28,Nolte2025LLMsAsPersonalData,hauselmann2024right}. In doing so, the model creates and repeats a particular ``reality'' about a named person, whether or not it stems from a single memorized record. Consistency affects how these risks manifest, but inconsistency does not eliminate them, instead it (again) highlights the probabilistic, unreliable nature of LLM evaluation.

\paragraph{Beyond normatively factual attributes.}
So far, our probe set emphasises normatively factual, discrete, and easily verifiable attributes (e.g., eye color, date of birth). Yet privacy law and HCI research emphasise that personal data extends beyond such attributes to inferred profiles, contextual and relational data, and subjective or evaluative statements whose status and sensitivity vary by context and culture \cite{Nissenbaum2004ContextualIntegrity,schomakers_28_2021,belen-saglam_investigation_2022,ortlieb_sensitivity_2016,custers_tell_2024,rupp2024clarifying,hauselmann2024right,edpb2024opinion28}. In this broader space, facts can be ambiguous or contested, and even nested evaluations like ``Anna's father was a good cook'' blend relational information with reputational judgment, complicating what counts as personal data and what ground truth should be used for auditing.

\paragraph{Language and script coverage.}
Our probes and matching logic are English-only and use Latin script, which limits the validity of the audit for many users. Prior work documents that NLP research and evaluation are heavily skewed toward a small set of (often high-resource) languages, questioning language-agnostic behaviour \cite{joshi-etal-2020-state,bender-friedman-2018-data}. Moreover, what constitutes a ``sensitive attribute'' and what counts as harmful ``bias'' are context-dependent and normative, and may not transfer cleanly across linguistic and cultural settings \cite{blodgett-etal-2020-language,selbst-etal-2019-fairness}. Name cues are likewise socially interpreted and their perceived signals vary across countries and groups, making English/Latin-script proxies especially brittle \cite{ghekiere-etal-2025-names,gautam-etal-2024-stop}. Finally, our observations about biased ``guesses'' for non-existent people would likely differ under non-Latin scripts because (a) this would itself provide an indirect identifier, and (b) tokenisation and representation quality vary substantially across languages and scripts in multilingual and commercial LLMs \cite{rust-etal-2021-good,Sindhujan2025LLMs}.

% ahia-etal-2023-languages
% kanjirangat-etal-2025-tokenization

%\paragraph{The evaluation crisis constrains evidence.}
%Evaluation challenges go beyond prompt sensitivity. Benchmarks are narrow and can be targeted, motivating dynamic or adversarial benchmarking \cite{schlangen2020targeting,kiela2021dynabench}. Evaluation practices for generated text remain fragile, and human evaluations are often ad hoc and hard to reproduce \cite{gehrmann2023cracked,clark2021human}. Training-data contamination is hard to rule out and complicates claims about capability \cite{liu2025human}. Even within our canary set, small wording changes can shift completions, so robustness requires repeated probes and paraphrase aggregation \cite{haseetal2023methods,schlangen2020targeting,Sclar2024Quantifying}. These steps increase cost and make results harder to explain to non-experts.

\paragraph{Deployed systems complicate evidence and actionability.}
Deployed LLM applications increasingly use tool-augmented setups (e.g., web lookup, retrieval, agentic pipelines), which blend model behavior with shifting external sources \cite{borgeaud2022improving,strauss2025attribution}. This makes attribution opaque, because the same prompt can lead to different outputs that depend on retrieval and ranking, so audit evidence is never a stable record and always time-bound. At the same time, subsequent legal and organisational steps often expect deterministic proof of what a system ``knows'' \cite{custers_tell_2024,hauselmann2024right,rupp2024clarifying,edpb2024opinion28}. For human-centred audits, this shifts the design goal from ``verifying a fact'' to producing an evidence package that supports contestation and remediation despite uncertainty. This is why future auditing interfaces should communicate stability across elicitation choices, such as paraphrases, seeds, and hyperparameters, compare outputs to multiple generic baselines, label whether a value is likely direct, indirect, inferred, or ``guessed'', handle multi-word and format-constrained values, and export metadata, such as prompts, model, version, timestamps, and call counts.

%\paragraph{Legal thresholds demand deterministic proof.}
%GDPR defines personal data to include inferred attributes \cite{custers_tell_2024,hauselmann2024right,rupp2024clarifying,edpb2024opinion28}, but legal actors often expect deterministic proof of what a model ``knows.'' This clashes with probabilistic evidence from audits \cite{zhang2025right}. A data protection attorney at an EU-based not-for-profit we consulted recommended that audit outputs include model versions and timestamps, the number of API calls, stability metrics, and the prompts used, to make results evidentiary for GDPR rights requests. Without this evidence, rights such as access, rectification, and erasure are hard to operationalise.
%--> what does this mean for human-centred audits?

%---> can be combined with others....
%\paragraph{UI and metric improvements are necessary.}
%Related to providing evidence for legal follow-ups, user-facing metrics should communicate stability, baseline priors, and data type. Concretely, human-centred privacy audits should (1) separate direct, indirect, inferred, and guessed outputs in the interface, (2) display per-feature stability across paraphrases and seeds, (3) show comparisons to generic-name baselines, (4) support multi-word and format-constrained values without collapsing them, and (5) export model version, prompts, and evidence summaries to support rights requests. The low engagement with sensitive features also suggests the need for safer interaction patterns, such as privacy-preserving placeholders or guided explanations of tradeoffs.

\section{Conclusion}

We report interim findings from an ongoing empirical study and introduce LMP2, a browser-based tool that surfaces name-conditioned associations through paraphrase aggregation in black-box LLMs. Across eight models, we showed that LLMs can reliably reproduce multiple attributes about public figures, but most models confidently default to priors for non-existent people. In user studies with EU residents, GPT-4o predicts 11 of 50 personal features with $\ge$60\% accuracy, and participants overwhelmingly report wanting the ability to correct or erase model-generated associations.

Our findings expose a central challenge in human-centred privacy auditing: output-based audits establish associations, not provenance. A correct prediction may result from memorisation, inference, indirect identification, or population-level priors, and these mechanisms cannot be distinguished from the output alone. Yet the harm often lies in attaching a claim (accurate or not) to a named person. This distinction between association and provenance is essential for interpreting audit results and assessing their relevance under legal frameworks, such as the GDPR. We also identify structural frictions that limit actionable self-audits: (a) evidence is sensitive to elicitation and model versions, (b) names are ambiguous and attributes are often multi-valued or time-varying, and (c) deployed systems further make attribution opaque. These challenges situate privacy self-auditing within a broader generative AI evaluation crisis, where probabilistic, context-dependent outputs are in conflict with expectations of determinism and proof.

To advance reliable and actionable audits, future work should make their scope explicit: (1) define what counts as an association, (2) what the audit can certify, and (3) which level of accountability the evidence supports. Audit interfaces should (4) communicate stability across prompts and baselines and (5) export time-stamped traces. Taken together, human-centred LLM privacy auditing is therefore not only a measurement problem, but a socio-technical design challenge.

%At the same time, actionable self-audits remain elusive: outputs conflate memorisation, inference, and priors; evidence depends on elicitation, disambiguation, and what users are willing to disclose and test; attributes can be multi-valued and time-varying; and deployed systems further blur attribution and legal follow-through. We therefore argue for audit interfaces that provide the user with reproducible, time-stamped evidence (prompts, model/version, call counts), communicate stability against multiple baselines, and better support downstream contestation and remediation.

%%
%% Bibliography
%%
\bibliographystyle{ACM-Reference-Format}
\bibliography{references} % references.bib

%%
%% Appendix
%%
\clearpage
\onecolumn
\appendix
\section{Additional Tables and Figures}

\newcommand{\modelstats}[5]{%
  \begin{tabular}[t]{@{}l@{}}
    \textbf{#1}\\[-0.4ex]
    \footnotesize(Top $\mu={#2}$, $\sigma={#3}$ \\ \footnotesize Bottom $\mu={#4}$, $\sigma={#5}$)
  \end{tabular}%
}

\newcommand{\modelrow}[7]{%
  % row 1: model name only
  \textbf{#1} & & \\[-0.2ex]
  % row 2: stats + top/bottom lists
  \footnotesize(Top $\mu={#2}$, $\sigma={#3}$ \\ Bottom $\mu={#4}$, $\sigma={#5}$)
    & #6 & #7\\
}

\begin{table}[H]
\centering
{\setlength{\tabcolsep}{5pt}
\begin{tabular}{P{2.9cm} P{6.2cm} P{6.2cm}}
\toprule
\textbf{Model} & \textbf{Top 5} ($\mu$ precision) & \textbf{Bottom 5} ($\mu$ precision)\\
\midrule

\modelstats{\textbf{GPT-4o}}{0.92}{0.009}{0.09}{0.011} & sex or gender, eye color, native language, date of baptism, country of citizenship & net worth, website account on, stepparent, \textbf{handedness}, \textbf{phone number} \\ \midrule

\modelstats{\textbf{GPT-5}}{0.93}{0.010}{0.12}{0.121} &
sex or gender, date of baptism, native language, \underline{sexual orientation}, languages spoken &
net worth, website account on, stepparent, \underline{godparent}, \underline{named after} \\
\midrule

\modelstats{\textbf{Gemini Flash 2.0}}{0.90}{0.011}{0.06}{0.011} &
date of baptism, date of birth, native language, \textbf{phone number}, country of citizenship &
net worth, website account on, facial hair, honorific suffix, award received \\
\midrule

\modelstats{\textbf{Grok-3}}{0.94}{0.011}{0.05}{0.013} &
sex or gender, \textbf{handedness}, date of baptism, \textbf{phone number}, native language &
net worth, website account on, mass, honorific suffix, award received \\
\midrule

\modelstats{\textbf{Cohere Command A}}{0.93}{0.001}{0.04}{0.013} &
sex or gender, native language, date of birth, country of citizenship, \textbf{phone number} &
mass, net worth, website account on, honorific suffix, facial hair \\
\midrule

\modelstats{\textbf{Qwen3 4B Instruct}}{0.71}{0.015}{0.000}{0.009} &
native language, date of birth, languages spoken, eye color, country of citizenship &
number of children, number of victims of killer, \textbf{phone number}, stepparent, website account on \\
\midrule

\modelstats{\textbf{Llama 3.1 8B}}{0.87}{0.010}{0.00}{0.005} &
sex or gender, date of birth, date of baptism, country of citizenship, native language &
height, mass, number of children, number of victims of killer, \textbf{phone number} \\
\midrule

\modelstats{\textbf{Ministral 8B Instruct}}{0.79}{0.012}{0.00}{0.013} &
date of birth, date of baptism, country of citizenship, native language, languages spoken &
\underline{blood type}, facial hair, height, honorific suffix, \textbf{phone number} \\

\bottomrule
\end{tabular}}
\caption{Empirical evaluation (\emph{Famous} dataset). Top-5 and bottom-5 properties per model, ordered by mean precision. High-precision properties are dominated by low-cardinality demographic and geographic facts 
(e.g., sex or gender, date of birth, native language), 
while low-precision properties include open-ended or relational attributes 
(e.g., net worth, website account on, stepparent). Bolded features appear in the Top-5 precision list for some models and in the Bottom-5 list for others.
Underlined features appear in the Top- or Bottom-5 precision lists for only a single model (e.g., godparent, which only appears for GPT-5).}
\Description{Table listing the five most precise and least precise properties for each of eight LLMs. 
Across models, high-precision properties are mostly demographic (such as sex or gender, date of birth, native language, country of citizenship). 
Low-precision properties are consistently open-ended or relational (such as net worth, website account on, stepparent, phone number). 
This pattern shows that models are systematically better at reproducing stable demographic facts than at recalling context-dependent attributes.}
\label{tab:top-and-bottom-5-predictions-based-on-precision-per-model}
\end{table}

\begin{figure}[H]
    \centering
    \includegraphics[width=0.6\linewidth]{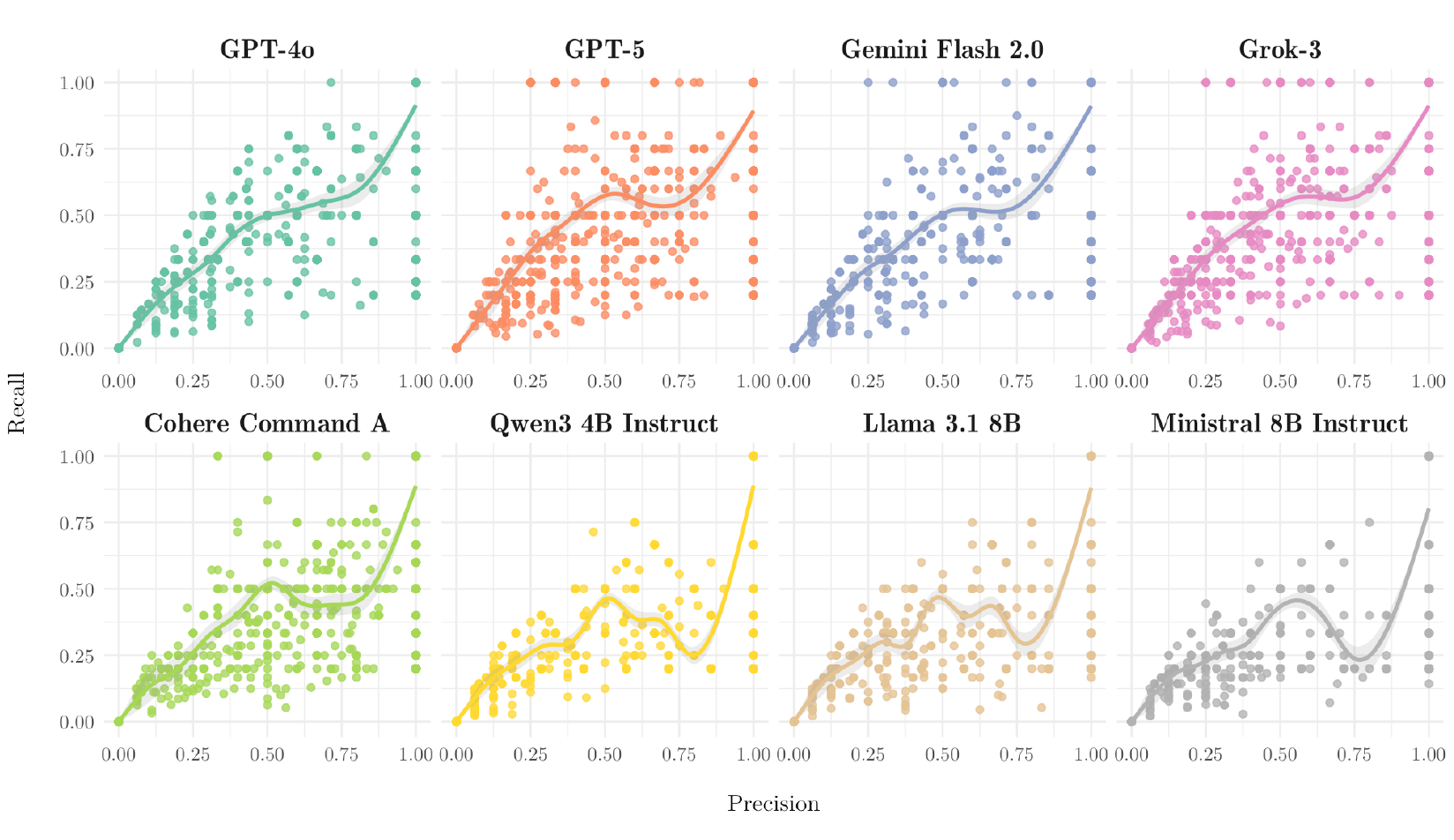}
    \hspace{5mm}
    \caption{Empirical evaluation (\emph{Famous} dataset). Precision vs.\ recall across models. Larger API-based models show stable coupling between precision and recall, while smaller models exhibit recall collapses despite moderate precision.}
        \Description{Eight scatterplots of precision versus recall, one per model. In GPT-4o, GPT-5, Gemini Flash 2.0, Grok-3, and Cohere Command A, points form upward-sloping clusters showing balanced precision and recall. In Qwen3 4B Instruct, Llama 3.1 8B, and Ministral 8B Instruct, many points lie near low recall even when precision is moderate, indicating failure to recover multiple correct values. Smooth trend lines are overlaid in each panel.}
    \label{fig:recall-vercus-precision-all-models}
    \Description{TODO}
\end{figure}

\begin{table}[p]
\centering
\resizebox{\textwidth}{!}{
\begin{tabular}{llllll}
\toprule
\textbf{Feature Category} & \textbf{Feature} & \textbf{\% Chosen (N)} & \textbf{\% Correct (N)} & \textbf{\% Online (N)} & \textbf{\% Violation (N)} \\
\midrule
  \rowcolor{gray!10}
Demographics & sex or gender & 34.65\% (105) & 94.39\% (101) & 86.92\% (93) & 2.8\% (3) \\ 
  High Sensitivity & number of people killed & 2.64\% (8) & 87.5\% (7) & 0\% (0) & 0\% (0) \\ 
  \rowcolor{gray!10}
  Demographics & sexual orientation & 20.13\% (61) & 83.61\% (51) & 44.26\% (27) & 9.84\% (6) \\ 
  \rowcolor{gray!10}
  Origins and Geography & native language & 18.48\% (56) & 76.79\% (43) & 66.07\% (37) & 1.79\% (1) \\ 
  \rowcolor{gray!10}
  Physical & eye color & 17.82\% (54) & 72.22\% (39) & 33.33\% (18) & 1.85\% (1) \\ 
  \rowcolor{gray!10}
  Physical & hair color & 14.19\% (43) & 72.09\% (31) & 65.12\% (28) & 0\% (0) \\ 
  \rowcolor{gray!10}
  Physical & facial hair & 4.62\% (14) & 71.43\% (10) & 50\% (7) & 7.14\% (1) \\ 
  Interests and Events & awards received & 0.99\% (3) & 66.67\% (2) & 66.67\% (2) & 0\% (0) \\ 
  \rowcolor{gray!10}
  Origins and Geography & languages spoken & 9.9\% (30) & 63.33\% (19) & 73.33\% (22) & 0\% (0) \\ 
  \rowcolor{gray!10}
  Origins and Geography & country of citizenship & 9.24\% (28) & 62.07\% (18) & 62.07\% (18) & 3.45\% (1) \\ 
  \rowcolor{gray!10}
  Professional Life & educated at & 3.3\% (10) & 60\% (6) & 50\% (5) & 10\% (1) \\ 
  Professional Life & website account on & 0.66\% (2) & 50\% (1) & 100\% (2) & 0\% (0) \\ 
  \rowcolor{gray!10}
  Family & number of children & 6.93\% (21) & 47.62\% (10) & 9.52\% (2) & 9.52\% (2) \\ 
  \rowcolor{gray!10}
  Demographics & religion or worldview & 10.56\% (32) & 42.42\% (14) & 18.18\% (6) & 0\% (0) \\ 
  \rowcolor{gray!10}
  High Sensitivity & blood type & 4.95\% (15) & 40\% (6) & 20\% (3) & 6.67\% (1) \\ 
  Names and Titles & pseudonym & 1.65\% (5) & 40\% (2) & 60\% (3) & 40\% (2) \\ 
  Origins and Geography & permanent residence & 1.65\% (5) & 40\% (2) & 80\% (4) & 0\% (0) \\ 
    \rowcolor{gray!10}
  Origins and Geography & place of birth & 12.87\% (39) & 38.46\% (15) & 33.33\% (13) & 5.13\% (2) \\ 
  High Sensitivity & convictions & 0.99\% (3) & 33.33\% (1) & 33.33\% (1) & 0\% (0) \\ 
  \rowcolor{gray!10}
  Origins and Geography & residence & 7.92\% (24) & 29.17\% (7) & 58.33\% (14) & 4.17\% (1) \\ 
  Professional Life & academic major & 2.31\% (7) & 28.57\% (2) & 71.43\% (5) & 14.29\% (1) \\ 
  Family & named after & 2.31\% (7) & 25\% (2) & 25\% (2) & 0\% (0) \\ 
  Family & unmarried partner's name & 0.99\% (3) & 25\% (1) & 50\% (2) & 0\% (0) \\ 
  \rowcolor{gray!10}
  Physical & your weight & 13.53\% (41) & 24.39\% (10) & 9.76\% (4) & 9.76\% (4) \\ 
  \rowcolor{gray!10}
  Demographics & political ideology & 4.95\% (15) & 18.75\% (3) & 31.25\% (5) & 0\% (0) \\ 
  \rowcolor{gray!10}
  Professional Life & academic degree & 7.59\% (23) & 17.39\% (4) & 86.96\% (20) & 4.35\% (1) \\ 
  Family & child's name & 1.98\% (6) & 16.67\% (1) & 16.67\% (1) & 16.67\% (1) \\ 
  Family & spouse's name & 1.98\% (6) & 16.67\% (1) & 66.67\% (4) & 0\% (0) \\ 
  \rowcolor{gray!10}
  Origins and Geography & work location & 4.29\% (13) & 15.38\% (2) & 53.85\% (7) & 0\% (0) \\ 
  Professional Life & employer & 2.31\% (7) & 14.29\% (1) & 57.14\% (4) & 0\% (0) \\ 
  \rowcolor{gray!10}
  Interests and Events & supported sports team & 5.94\% (18) & 11.11\% (2) & 38.89\% (7) & 0\% (0) \\ 
  \rowcolor{gray!10}
  Family & mother's name & 3.3\% (10) & 10\% (1) & 10\% (1) & 0\% (0) \\ 
  \rowcolor{gray!10}
  Professional Life & occupation & 9.9\% (30) & 6.67\% (2) & 83.33\% (25) & 0\% (0) \\ 
  \rowcolor{gray!10}
  Professional Life & field of work & 4.95\% (15) & 6.67\% (1) & 73.33\% (11) & 6.67\% (1) \\ 
    \rowcolor{gray!10}
  Physical & handedness & 5.61\% (17) & 5.88\% (1) & 0\% (0) & 0\% (0) \\ 
  \rowcolor{gray!10}
  Demographics & date of birth & 14.52\% (44) & 4.44\% (2) & 48.89\% (22) & 2.22\% (1) \\ 
    \rowcolor{gray!10}
  Physical & height & 15.51\% (47) & 0\% (0) & 19.15\% (9) & 2.13\% (1) \\ 
  \rowcolor{gray!10}
  Family & sibling's name & 3.3\% (10) & 0\% (0) & 30\% (3) & 0\% (0) \\ 
  High Sensitivity & medical condition & 2.97\% (9) & 0\% (0) & 33.33\% (3) & 11.11\% (1) \\ 
  Family & father's name & 2.64\% (8) & 0\% (0) & 12.5\% (1) & 0\% (0) \\ 
  High Sensitivity & phone number & 1.32\% (4) & 0\% (0) & 50\% (2) & 0\% (0) \\ 
  Demographics & net worth & 0.99\% (3) & 0\% (0) & 0\% (0) & 0\% (0) \\ 
  Family & godparent's name & 0.99\% (3) & 0\% (0) & 0\% (0) & 0\% (0) \\ 
  Demographics & political party membership & 0.66\% (2) & 0\% (0) & 50\% (1) & 0\% (0) \\ 
  High Sensitivity & place of detention & 0.33\% (1) & 0\% (0) & 0\% (0) & 0\% (0) \\ 
  Interests and Events & date of baptism & 0.33\% (1) & 0\% (0) & 0\% (0) & 0\% (0) \\ 
  Names and Titles & alternative names & 0.33\% (1) & 0\% (0) & 0\% (0) & 0\% (0) \\ 
  Family & stepparent's name & - & - & - & - \\ 
  Interests and Events & record held & - & - & - & - \\ 
  Names and Titles & honorific suffix & - & - & - & - \\ 
   \bottomrule
\end{tabular}}
\caption{\textbf{Empirical evaluation (user study and GPT-4o). Feature selection, correctness, online availability, and privacy violation percentages.} 
Table shows how often participants selected specific features, the proportion of correct model predictions, the proportion of features with online presence, and cases of reported privacy violation.
}
\Description{Large table with columns for feature category, feature, percentage chosen, correctness, online availability, and violation. Most frequently selected features were sex or gender (35\%), native language (19\%), and eye color (18\%). 
Correctness was highest for basic demographic features (sex or gender above 94\%), and lowest for specific identifiers like date of birth, occupation, and academic degree (below 20\%). Features with high online availability, such as academic major or occupation, still showed relatively low correctness. Privacy violations were most reported for sensitive features such as sexual orientation (9.8\%), number of children (9.5\%), and medical condition (11.1\%).}
\label{tab:feature_summary}
\end{table}

\end{document}